\documentclass[aps,prl,twocolumn,showpacs,superscriptaddress,groupedaddress,nofootinbib]{revtex4-2}  
\usepackage{cancel}
\usepackage{graphicx} 
\usepackage[T1]{fontenc}
\usepackage{lmodern} 
\usepackage{graphicx}  
\usepackage{dcolumn}   
\usepackage{bm}        
\usepackage{amssymb}   
\usepackage{amsmath}
\usepackage{bbold}
\usepackage{array}
\usepackage{float}
\usepackage{makecell}
\usepackage{braket}
\usepackage{longtable}
\usepackage{supertabular,booktabs}
\usepackage{soul}

\usepackage{titlesec} 
\usepackage[colorlinks,citecolor=blue,urlcolor=blue,hypertexnames=true]{hyperref}
\usepackage{subfigure}

\usepackage[usenames,dvipsnames,svgnames,table]{xcolor} 

\newcommand{\be}{\begin{equation}}
\newcommand{\ee}{\end{equation}}
\newcommand{\bea}{\begin{eqnarray}}
\newcommand{\eea}{\end{eqnarray}}
\newcommand{\bml}{\begin{subequations}}
\newcommand{\eml}{\end{subequations}}
\newcommand{\bfig}{\begin{figure}}
\newcommand{\efig}{\end{figure}}

\newcommand{\slashed}{\hspace{-1.1ex}/}

\newcommand{\bmat}{\begin{pmatrix}}
\newcommand{\emat}{\end{pmatrix}}
\usepackage{graphicx, slashed,booktabs, color, multirow, float,
amsfonts, bbold, mathtools, sidecap, tikz, bm,enumitem}
\usepackage{multirow}
\usepackage{bbding}
\usepackage{titlesec}
\usepackage{hyperref}
\usepackage{wasysym}
\usepackage{amssymb}
\usepackage{pifont}

\usepackage[dvipsnames, usenames]{xcolor}

\definecolor{linkcolor}{rgb}{0.55, 0.13, .32}

\definecolor{oucrimsonred}{rgb}{0.6, 0.0, 0.0}
\definecolor{persianblue}{rgb}{0.11, 0.22, 0.73}
\definecolor{forestgreen}{rgb}{0.13,0.35,0.13}
\definecolor{lightgray}{rgb}{0.83, 0.83, 0.83}
 \hypersetup{colorlinks, citecolor=oucrimsonred, linkcolor=persianblue, urlcolor=oucrimsonred}
\definecolor{cornellred}{rgb}{0.7, 0.11, 0.11}
\definecolor{navyblue}{rgb}{0.0, 0.0, 0.5}
\definecolor{amethyst}{rgb}{0.6, 0.4, 0.8}
\definecolor{yellow}{rgb}{1.0, 1.0, 0.0}
\definecolor{firebrick}{rgb}{0.7, 0.13, 0.13}
\definecolor{tangerineyellow}{rgb}{1.0, 0.8, 0.0}
\definecolor{deepfuchsia}{rgb}{0.76, 0.33, 0.76}
\definecolor{amber}{rgb}{1.0, 0.75, 0.0}
\definecolor{VioletRed4}{rgb}{0.55, 0.13, .32}
\definecolor{indiagreen}{rgb}{0.07, 0.53, 0.03}
\definecolor{VioletRed4}{rgb}{0.55, 0.13, .32}

\usepackage{hyperref}
\usepackage{graphics, appendix, afterpage, makecell} 
\usepackage{bbold}
\usepackage{tikz}
\usepackage{adjustbox}

\usepackage{tcolorbox}

\definecolor{oucrimsonred}{rgb}{0.6, 0.0, 0.0}
\definecolor{persianblue}{rgb}{0.11, 0.22, 0.73}
\definecolor{forestgreen}{rgb}{0.13,0.35,0.13}
\definecolor{lightgray}{rgb}{0.83, 0.83, 0.83}
 \hypersetup{colorlinks, citecolor=oucrimsonred, linkcolor=persianblue, urlcolor=oucrimsonred}
\definecolor{cornellred}{rgb}{0.7, 0.11, 0.11}
\definecolor{navyblue}{rgb}{0.0, 0.0, 0.5}
\definecolor{amethyst}{rgb}{0.6, 0.4, 0.8}
\definecolor{yellow}{rgb}{1.0, 1.0, 0.0}
\definecolor{firebrick}{rgb}{0.7, 0.13, 0.13}
\definecolor{tangerineyellow}{rgb}{1.0, 0.8, 0.0}
\definecolor{deepfuchsia}{rgb}{0.76, 0.33, 0.76}
\definecolor{amber}{rgb}{1.0, 0.75, 0.0}
\definecolor{VioletRed4}{rgb}{0.55, 0.13, .32}
\definecolor{indiagreen}{rgb}{0.07, 0.53, 0.03}
\definecolor{VioletRed4}{rgb}{0.55, 0.13, .32}

\definecolor{oucrimsonred}{rgb}{0.6, 0.0, 0.0}
\newcommand\vertarrowbox[3][6ex]{%
  \begin{array}[t]{@{}c@{}} #2 \\
  \left\uparrow\vcenter{\hrule height #1}\right.\kern-\nulldelimiterspace\\
  \makebox[0pt]{\scriptsize#3}
  \end{array}%
}

\definecolor{mtcolor}{rgb}{.8,.3,.1}

\definecolor{violachiaro}{rgb}{1,0.6,1}

\definecolor{gbcolor}{rgb}{.43,.22,.12}
 
\definecolor{gbcolor2}{rgb}{.9,.2,.6}
\definecolor{gbcolor3}{rgb}{.3,.2,.6}

\definecolor{verdechiaro}{rgb}{0.6,1,0.6}
\definecolor{giallochiaro}{rgb}{1,1,0.6}
\definecolor{bluscuro}{rgb}{0.15, 0.2, 0.9}
\definecolor{verdes}{rgb}{0.1, 0.5, 0.1}%
\definecolor{tangerineyellow}{rgb}{1.0, 0.8, 0.0}
\definecolor{smokyblack}{rgb}{0.06, 0.05, 0.03}

\definecolor{americanrose}{rgb}{1.0, 0.01, 0.24}
\definecolor{cobalt}{rgb}{0.0, 0.28, 0.67}
\definecolor{brandeisblue}{rgb}{0.0, 0.44, 1.0}
\definecolor{mycolor}{rgb}{0.0, 0.0, 0.5}
\definecolor{oxfordblue}{rgb}{0.0, 0.13, 0.28}
\definecolor{azure}{rgb}{0.0, 0.5, 1.0}
\definecolor{turquoiseblue}{rgb}{0.0, 1.0, 0.94}
\newtcolorbox{mynewbox}[1]{colback=white!5!white,colframe=azure!75!black,fonttitle=\bfseries,title=#1}
\newtcolorbox{mybox}{colback=mycolor!5!white,colframe=azure!75!black}
\newtcolorbox{mynamedbox}[1]{colback=mycolor!5!white,colframe=azure!75!black,title=#1}
\definecolor{venetianred}{rgb}{0.78, 0.03, 0.08}
\newtcolorbox{mynamedbox1}[1]{colback=venetianred!5!white,colframe=venetianred!80!black,title=#1}
\newtcolorbox{mynamedbox2}[1]{colback=azure!5!white,colframe=azure!80!black,title=#1}

\definecolor{rossocorsa}{rgb}{0.83, 0.0, 0.0}

\tikzset{->-/.style={decoration={
  markings,
  mark=at position #1 with {\arrow{>}}},postaction={decorate}}}
\tikzset{-<-/.style={decoration={
  markings,
  mark=at position #1 with {\arrow{<}}},postaction={decorate}}} 

\def\be{\begin{equation}}
\def\ee{\end{equation}}
\def\ba{\begin{eqnarray}}
\def\ea{\end{eqnarray}}

\def\L*{{\cal L}_*}
\def\L{\mathcal{L}}
\def\({\left(}
\def\){\right)}

\def\<{\langle}
\def\>{\rangle}

 \def\neq {\not\equiv}

\def\cs2{c_{s}^{2}}

 \def\be   {\begin{equation}}   \def\ee   {\end{equation}}
 \def\ba   {\begin{array}}      \def\ea   {\end{array}}
 \def\bea  {\begin{eqnarray}}   \def\eea  {\end{eqnarray}}
 \def\bean {\begin{eqnarray*}}  \def\eean {\end{eqnarray*}}

\titleclass{\subsubsubsection}{straight}[\subsection]

\newcounter{subsubsubsection}[subsubsection]
\renewcommand\thesubsubsubsection{\thesubsubsection.\arabic{subsubsubsection}}

\titleformat{\subsubsubsection}
  {\normalfont\normalsize\bfseries}{\thesubsubsubsection}{1em}{}
\titlespacing*{\subsubsubsection}
{0pt}{3.25ex plus 1ex minus .2ex}{1.5ex plus .2ex}

\makeatletter
\renewcommand\paragraph{\@startsection{paragraph}{5}{\z@}%
  {3.25ex \@plus1ex \@minus.2ex}%
  {-1em}%
  {\normalfont\normalsize\bfseries}}
\renewcommand\subparagraph{\@startsection{subparagraph}{6}{\parindent}%
  {3.25ex \@plus1ex \@minus .2ex}%
  {-1em}%
  {\normalfont\normalsize\bfseries}}
\def\toclevel@subsubsubsection{4}
\def\toclevel@paragraph{5}
\def\toclevel@paragraph{6}
\def\l@subsubsubsection{\@dottedtocline{4}{7em}{4em}}
\def\l@paragraph{\@dottedtocline{5}{10em}{5em}}
\def\l@subparagraph{\@dottedtocline{6}{14em}{6em}}
\makeatother

\usepackage{titlesec} 
\usepackage[colorlinks,citecolor=blue,urlcolor=blue,hypertexnames=true]{hyperref}
\usepackage{subfigure}

\usepackage[usenames,dvipsnames,svgnames,table]{xcolor}

\usepackage{tikzsymbols}
\usepackage{natbib}
\usepackage{float}

\usepackage{tikz,xcolor,hyperref}

\usepackage{slashed}

\definecolor{lime}{HTML}{A6CE39}
\DeclareRobustCommand{\orcidicon}{
	\begin{tikzpicture}
	\draw[lime, fill=lime] (0,0) 
	circle [radius=0.2] 
	node[white] {{\fontfamily{qag}\selectfont \tiny ID}};
	\draw[white, fill=white] (-0.0625,0.095) 
	circle [radius=0.007];
	\end{tikzpicture}
	\hspace{-2mm}
}

\usepackage[usenames,dvipsnames,svgnames,table]{xcolor}  
\usepackage{tikzsymbols}
\usepackage{natbib}
\usepackage{float}

\usepackage{tikz,xcolor,hyperref}

\usepackage{slashed}

\definecolor{lime}{HTML}{A6CE39}
\DeclareRobustCommand{\orcidicon}{
	\begin{tikzpicture}
	\draw[lime, fill=lime] (0,0) 
	circle [radius=0.2] 
	node[white] {{\fontfamily{qag}\selectfont \tiny ID}};
	\draw[white, fill=white] (-0.0625,0.095) 
	circle [radius=0.007];
	\end{tikzpicture}
	\hspace{-2mm}
}

\foreach \x in {A, ..., Z}{\expandafter\xdef\csname orcid\x\endcsname{\noexpand\href{https://orcid.org/\csname orcidauthor\x\endcsname}
			{\noexpand\orcidicon}}
}
 
\usepackage{bbold}
\usepackage{tikz}
\usepackage{adjustbox}
\usepackage{tcolorbox}
\usepackage{enumitem}
\usepackage{amsfonts}

\setlist[itemize,1]{label=$\times$}
\setlist[itemize,2]{label=$\checkmark$}
\setlist[itemize,3]{label=$\diamond$}
\setlist[itemize,4]{label=$\bullet$}

\begin{document}
\title{\Large \textcolor{Sepia}{Gravitational Wave Signatures of Periodic Motion near Higher-Derivative Einstein-\AE ther Black Holes
}}
\author{\large Sayantan Choudhury\orcidA{}${}^{1}$}
\email{sayanphysicsisi@gmail.com, \\
schoudhury@fuw.edu.pl,\\
sayantan.choudhury@nanograv.org (Corresponding author)} 
\author{\large Md Khalid Hossain\orcidF{}${}^{2}$}
\email{mdkhalidhossain600@gmail.com}
\author{\large Gulnur Bauyrzhan \orcidC{}${}^{3}$}
\email{baurzhan.g.b@gmail.com} 
\author{\large Koblandy Yerzhanov\orcidB{}${}^{3}$}
\email{yerzhanovkk@gmail.com}

\affiliation{ ${}^{1}$Institute of Theoretical Physics, Faculty of Physics,\\
University of Warsaw, ul. Pasteura 5, 02-093 Warsaw, Poland,}
\affiliation{ ${}^{2}$Department of Mathematics, Jadavpur University, Kolkata 700032, West Bengal, India,}
\affiliation{ ${}^{3}$Center for Theoretical Physics,  L.N. Gumilyov Eurasian National University, Astana 010008, Kazakhstan.}







\begin{abstract}

Higher-derivative modifications of general relativity are generically expected from effective field theory approaches to quantum gravity, and they arise naturally in Lorentz-violating theories such as Einstein-Æther gravity. In this work, we investigate black hole spacetimes within Einstein-Æther theory supplemented by quadratic curvature corrections, including terms proportional to $R^2$, $R_{\mu\nu} R^{\mu\nu}$, and $R_{\mu\nu\lambda\rho} R^{\mu\nu\lambda\rho}$. We derive the corrected static, spherically symmetric metric perturbatively and examine its effects on the geodesic structure and gravitational wave emission. In particular, we analyze periodic timelike orbits in this background and compute the associated tensor-mode gravitational waveforms using the quadrupole approximation. Our results demonstrate that even small higher-derivative corrections can induce distinguishable shifts in the orbital dynamics and imprint characteristic phase modulations and harmonic deformations in the gravitational wave signal. These effects modify the frequency spectrum and amplitude envelope of $h_{+}$ and $h_{\times}$ in a manner sensitive to the coupling constants $\alpha$, $\beta$, and $\gamma$, and the Æther parameter $c_{13}$. The resulting signatures provide a potential observational window into ultraviolet deviations from general relativity and Lorentz symmetry in the strong-field regime.


\end{abstract}

\maketitle

\section{Introduction}

The primary inspiration for this work comes from \cite{x} that discusses gravitational radiation emitted from periodic orbits around Einstein-Æther black holes.

The direct detection of gravitational waves by the LIGO and Virgo collaborations \cite{r1,r2,r3,r4} has transformed our ability to probe the strong-field, dynamical regime of gravity. While general relativity (GR) continues to pass observational tests with remarkable accuracy, numerous theoretical and phenomenological motivations suggest that GR may require modification at high energies. Effective field theory (EFT) approaches to quantum gravity predict that the Einstein-Hilbert action should be supplemented by higher-curvature corrections involving terms like $R^2$, $R_{\mu\nu} R^{\mu\nu}$, and $R_{\mu\nu\lambda\rho} R^{\mu\nu\lambda\rho}$ \cite{Stelle1977, Buchbinder1992, Woodard2007}. These terms arise in the low-energy limit of various quantum gravity proposals, including string theory and loop quantum gravity, and are expected to become relevant in the vicinity of compact objects.

Simultaneously, theories that break local Lorentz invariance—such as Einstein-Æther theory \cite{r5,r6,r7}—provide a consistent framework in which the gravitational field couples to a dynamical, unit-norm timelike vector field. This theory modifies the structure of gravitational radiation and black hole solutions while preserving diffeomorphism invariance. Linearized perturbation theory reveals up to five or six gravitational wave polarizations depending on the choice of coupling constants \cite{r7a}. However, for a specific region in parameter space where $c_{14} = 0$, only the two standard tensor modes of GR propagate, making the theory compatible with current gravitational wave observations \cite{r7a,r8,r9}.

In a series of papers \cite{r13,r14,r15,r16,r17,r18,r19,r20,r21}, Levin et al. previously characterized the anatomy of periodic, zoom-whirl orbits \cite{r22}. These descriptions of periodic orbits have recently been used in alternative gravity theories or other speculative spacetime contexts \cite{r23,r24,r25,r26,r27,r28,r29,r30,r31,r32}.

The waveform signal from compact binaries offers a powerful tool to test such deviations from GR. Gravitational wave emission in Lorentz-violating theories and higher-curvature gravity has been studied both analytically and numerically \cite{Barausse2011, Blas2011, Stein2014}. Of particular importance are extreme-mass-ratio inspirals (EMRIs), where a compact stellar-mass object orbits a supermassive black hole over thousands of cycles before merging. These systems, expected to be detected by future space-based interferometers such as LISA \cite{AmaroSeoane2017}, are sensitive to subtle features in the underlying gravitational theory.

In this work, we investigate the combined effects of higher-curvature corrections and æther field contributions on black hole spacetimes and gravitational radiation. We construct a perturbative static, spherically symmetric black hole solution in Einstein-Æther gravity, including curvature-square terms. We then analyze the structure of periodic orbits, with special focus on zoom-whirl trajectories, and compute the associated gravitational waveforms using the quadrupole approximation. Our analysis shows that even minute deviations from GR can lead to measurable phase shifts and harmonic distortions in the waveform, potentially distinguishable from standard predictions.

This study highlights how gravitational wave astronomy can serve as a precision probe of ultraviolet modifications to gravity and Lorentz symmetry breaking in the strong-field regime \cite{r10,r11,r12}.

\section{Black Holes in Higher-Derivative Einstein-\AE ther Theory}
\label{Important}
The aether field, a dynamical time-like unit vector field $u^\mu$that specifies a preferred local rest frame, is a component of Einstein- \AE ther theory, a Lorentz-violating modification of general relativity. The theory openly breaks local Lorentz symmetry while maintaining diffeomorphism invariance. The action describes its gravitational sector.

We consider the  action for higher derivative Einstein- \AE ther theory  :
\begin{widetext}
    \begin{eqnarray}\label{e1}
S = \frac{1}{16\pi G_{ae}} \int d^4x \sqrt{-g} \bigg( R + \mathcal{L}_{ae} + \alpha R^2 + \beta R_{\mu\nu} R^{\mu\nu} + \gamma R_{\mu\nu\lambda\rho} R^{\mu\nu\lambda\rho} \bigg) + \int d^4x \sqrt{-g} \, \mathcal{L}_M  
\end{eqnarray}
\end{widetext}

The variation of the action with respect to the metric $g^{\mu\nu}$ gives:
\begin{multline}
\delta S = \frac{1}{16\pi G_{ae}} \int d^4x \sqrt{-g}  \delta R + \delta \mathcal{L}_{ae} + \alpha \delta R^2 + \beta \delta(R_{\mu\nu} R^{\mu\nu}) \\ + \gamma \delta(R_{\mu\nu\lambda\rho} R^{\mu\nu\lambda\rho}) + \delta S_M
\end{multline}

The resulting field equations are:
\begin{equation}
G_{\mu\nu} + H^{ae}_{\mu\nu} + 16\pi G_{ae} \, T_{\mu\nu} = \alpha H^{R^2}_{\mu\nu} + \beta H^{R_{\mu\nu}^2}_{\mu\nu} + \gamma H^{R_{\mu\nu\lambda\rho}^2}_{\mu\nu}
\end{equation}

Where the individual higher-order contributions are:
\begin{widetext}
\begin{align}
H^{R^2}_{\mu\nu} &= 2 R R_{\mu\nu} - \frac{1}{2} R^2 g_{\mu\nu} - 2 \nabla_\mu \nabla_\nu R + 2 g_{\mu\nu} \Box R \\
H^{R_{\mu\nu}^2}_{\mu\nu} &= 2 R_{\mu\rho} R_\nu^{\ \rho} - \frac{1}{2} R_{\alpha\beta} R^{\alpha\beta} g_{\mu\nu} - \Box R_{\mu\nu} - \nabla_\mu \nabla_\nu R \notag \\ &+ g_{\mu\nu} \nabla_\alpha \nabla_\beta R^{\alpha\beta} \\
H^{R_{\mu\nu\lambda\rho}^2}_{\mu\nu} &= 2 R_{\mu\alpha\beta\gamma} R_\nu^{\ \alpha\beta\gamma} - \frac{1}{2} g_{\mu\nu} R_{\alpha\beta\gamma\delta} R^{\alpha\beta\gamma\delta} - 4 \nabla^\alpha \nabla^\beta R_{\mu\alpha\nu\beta}
\end{align}
\end{widetext}
The stress-energy tensor is defined as:
\begin{equation}
T_{\mu\nu} = -\frac{2}{\sqrt{-g}} \frac{\delta(\sqrt{-g} \, \mathcal{L}_M)}{\delta g^{\mu\nu}}
\end{equation}

The variation of the aether term $\mathcal{L}_{ae}$ depends on the specific formulation (e.g., Einstein-aether theory), and contributes a tensor:
\begin{equation}
H^{ae}_{\mu\nu} = -\frac{1}{\sqrt{-g}} \frac{\delta (\sqrt{-g} \mathcal{L}_{ae})}{\delta g^{\mu\nu}}
\end{equation}

\vspace{1em}
\noindent
This completes the derivation of the field equations for a theory with quadratic curvature corrections.

The standard Bianchi identity in General Relativity is
\begin{equation}
\nabla^\mu G_{\mu\nu} = 0 \,,
\end{equation}
which leads to local conservation of the stress-energy tensor:
\begin{equation}
\nabla^\mu T_{\mu\nu} = 0 \,.
\end{equation}

In the presence of higher-order curvature corrections, the total field equations take the form:
\begin{equation}
G_{\mu\nu} + H^{(ae)}_{\mu\nu} - \alpha H^{(R^2)}_{\mu\nu} - \beta H^{(R_{\mu\nu}^2)}_{\mu\nu} - \gamma H^{(R_{\mu\nu\lambda\rho}^2)}_{\mu\nu} = 8\pi G_{ae} \, T_{\mu\nu} \,.
\end{equation}

Taking the covariant divergence yields the generalized Bianchi identity:
\begin{widetext}
\begin{eqnarray}
 \nabla^\mu T_{\mu\nu} = \frac{1}{8\pi G_{ae}} \bigg[ \nabla^\mu H^{(ae)}_{\mu\nu} - \alpha \nabla^\mu H^{(R^2)}_{\mu\nu} - \beta \nabla^\mu H^{(R_{\mu\nu}^2)}_{\mu\nu} - \gamma \nabla^\mu H^{(R_{\mu\nu\lambda\rho}^2)}_{\mu\nu} \bigg] \,.   \end{eqnarray}
\end{widetext}

\section{Static and Spherically Symmetric Higher-Derivative Einstein-\AE ther Black Holes}

We now consider static and spherically symmetric black hole solutions in the Einstein-Æther theory with higher-derivative curvature corrections. The general form of the metric in Schwarzschild-like coordinates is taken as
\begin{equation}
ds^2 = -e(r)\, dt^2 + \frac{dr^2}{e(r)} + r^2 (d\theta^2 + \sin^2\theta\, d\varphi^2),
\end{equation}
where the metric function $e(r)$ incorporates both the Einstein-\AE ther and higher-curvature modifications.

Motivated by perturbative corrections, we expand $e(r)$ as
\begin{equation}
e(r) = 1 - \frac{2M}{r} + \delta e_{\ae}(r) + \delta e_{\text{HD}}(r),
\end{equation}
where $\delta e_{\ae} (r)$represents the correction from the \AE ther field, and  $\delta e_{\text{HD}}(r)$ accounts for contributions from higher-derivative curvature invariants. Specifically, for the first class of Einstein-\AE ther black holes (with $c_{14} = 0 $), the leading-order corrections are found to be
\begin{equation}
\delta e_{\ae}(r) = -\lambda_0 \left( \frac{2M}{r} \right)^4, \quad \lambda_0 \equiv \frac{27 c_{13}}{256(1 - c_{13})}.
\end{equation}

Incorporating the quadratic curvature corrections in the action eq.(\ref{e1})

the additional correction to $e(r)$ is given by
\begin{equation}
\delta e_{\text{HD}}(r) = \frac{\alpha_1}{r^2} + \alpha_2 \frac{\log r}{r^2},
\end{equation}
where the coefficients $\alpha_1$ and $\alpha_2$ depend on the coupling constants $ \alpha, \beta, \gamma $ as
\begin{align}
\alpha_1 &= -\frac{16\pi G_{ae}}{3} (3\alpha + \beta + 2\gamma) M, \\
\alpha_2 &= \frac{16\pi G_{ae}}{3} (\beta + 4\gamma) M.
\end{align}

Thus, the total corrected metric function up to $\mathcal{O}(1/r^4)$  becomes
\begin{widetext}
\begin{equation}
e(r) = 1 - \frac{2M}{r} + \frac{\alpha_1}{r^2} + \alpha_2 \frac{\log r}{r^2} - \lambda_0 \left( \frac{2M}{r} \right)^4 + \mathcal{O}\left( \frac{1}{r^5} \right).
\end{equation}
\end{widetext}

The corrections induced by the æther field and higher curvature terms play an important role in modifying the near-horizon structure of the black hole spacetime and can influence observable phenomena such as light deflection, perihelion precession, and gravitational wave emission.

\section{ Periodic Orbits in Higher-Derivative Einstein-\AE ther Theory}

In this section, we study the periodic timelike geodesics in the black hole spacetime modified by both the Einstein-Æther field and higher-derivative curvature terms. We consider the metric of the form:
\begin{equation}
ds^2 = -e(r) dt^2 + \frac{dr^2}{e(r)} + r^2(d\theta^2 + \sin^2\theta\, d\varphi^2),
\end{equation}
with the corrected metric function up to $\mathcal{O}(1/r^4)$ given by
\begin{equation}
e(r) = 1 - \frac{2M}{r} + \frac{\alpha_1}{r^2} + \alpha_2 \frac{\log r}{r^2} - \lambda_0 \left( \frac{2M}{r} \right)^4,
\end{equation}
where $\alpha_1$ and $\alpha_2$ are the higher-curvature coefficients, and $\lambda_0 = \frac{27 c_{13}}{256(1 - c_{13})}$ arises from Einstein-Æther corrections.

We consider the motion of a test particle described by the Lagrangian
\begin{equation}
\mathcal{L} = \frac{1}{2} g_{\mu\nu} \frac{dx^\mu}{d\lambda} \frac{dx^\nu}{d\lambda},
\end{equation}
where $\lambda$ is an affine parameter along the worldline. The conserved quantities associated with the Killing vectors $\partial_t$ and $\partial_\varphi$ are the energy $E$ and angular momentum $L$:
\begin{align}
p_t &= g_{tt} \dot{t} = -E, \\
p_\varphi &= g_{\varphi\varphi} \dot{\varphi} = L.
\end{align}

Restricting the motion to the equatorial plane ($\theta = \pi/2$, $\dot{\theta} = 0$), the normalization condition for timelike geodesics leads to:
\begin{equation}
\dot{r}^2 = E^2 - V_\text{eff}(r),
\end{equation}
where the effective potential is
\begin{equation}
V_\text{eff}(r) = \left(1 + \frac{L^2}{r^2} \right) e(r).
\end{equation}

Using the substitution $r = 1/x$, the radial equation becomes
\begin{equation}
\left( \frac{d\varphi}{dx} \right)^2 = \frac{L^2}{P(x)},
\end{equation}
where $P(x)$ is a polynomial depending on $x$ via the metric function $e(1/x)$. For the higher-derivative Einstein-Æther solution, this results in a sixth-order polynomial:
\begin{widetext}
    \begin{eqnarray}
  P(x) = E^2 - \left(1 + L^2 x^2 \right) \Bigg[ 1 - 2Mx + \alpha_1 x^2 + \alpha_2 x^2 \log\left(\frac{1}{x}\right)  - 16 M^4 \lambda_0 x^4 \Bigg].
  \end{eqnarray}
\end{widetext}

To analyze periodic orbits, we compute the ratio between the angular and radial frequencies:
\begin{equation}
q \equiv \frac{\omega_\varphi}{\omega_r} - 1 = \frac{1}{\pi} \int_{x_1}^{x_2} \frac{L}{\sqrt{P(x)}} dx - 1,
\end{equation}
where $x_1$ and $x_2$ are the turning points of the motion. The classification of periodic orbits is characterized by the triplet $(z, w, v)$, corresponding to the zoom, whirl, and vertex behavior of the orbit, respectively:
\begin{equation}
q = w + \frac{v}{z}.
\end{equation}

By tuning $(z, w, v)$ and the energy $E$, we generate families of closed periodic orbits. These orbits exhibit rich structures influenced by both the æther field and higher-derivative terms. Their topology can be visualized in polar coordinates and compared across different values of the parameters $\alpha$, $\beta$, $\gamma$, and $c_{13}$.

\bigskip

In the following section, we analyze the gravitational radiation associated with these orbits and investigate how the higher-derivative and æther corrections influence the waveform morphology.

\section{Gravitational Radiation in Higher-Derivative Einstein-\AE ther Theory}

An initial investigation of the gravitational radiation released by the periodic orbits of a test particle circling supermassive higher derivative Einstein-Æther black holes is presented in this section.

\subsection{ Polarizations of Gravitational Waves in Einstein-\AE ther Theory}

In Einstein-\AE ther theory, the presence of a unit timelike vector field $u^\mu$ leads to up to six independent polarization modes for gravitational waves. However, in the case where $c_{14} = 0$, all extra polarization modes vanish, and only the two tensorial modes, $h_+$ and $h_\times$, remain. This is consistent with current LIGO/Virgo/KAGRA observations that find no significant deviation from tensor polarizations.

We consider perturbations $h_{\mu\nu}$ on the background metric:
\begin{equation}
g_{\mu\nu} = \eta_{\mu\nu} + h_{\mu\nu}, \quad |h_{\mu\nu}| \ll 1.
\end{equation}

Working in the TT (transverse-traceless) gauge and Cartesian coordinates $(X,Y,Z)$, we define the polarization tensors:
\begin{align}
e^{+}_{ij} &= e^X_i e^X_j - e^Y_i e^Y_j, \\
e^{\times}_{ij} &= e^X_i e^Y_j + e^Y_i e^X_j,
\end{align}
where $e^X_i$, $e^Y_i$ are unit vectors perpendicular to the direction of wave propagation.

The gravitational wave polarizations are then given by:
\begin{align}
h_+(t) &= \frac{G_{ae}}{D_L} \ddot{Q}_{ij}(t) e^+_{ij}, \\
h_\times(t) &= \frac{G_{ae}}{D_L} \ddot{Q}_{ij}(t) e^\times_{ij},
\end{align}
where $Q_{ij}$ is the reduced quadrupole moment of the particle motion:
\begin{equation}
Q_{ij} = m \left(x_i x_j - \frac{1}{3} \delta_{ij} r^2\right).
\end{equation}

\subsection{ Gravitational Waves Emitted from Periodic Orbits in Einstein-Æther Theory}

We model a small mass $m$ orbiting a static black hole of mass $M$ in the higher-derivative Einstein-Æther background. The metric function is:
\begin{equation}
e(r) = 1 - \frac{2M}{r} + \frac{\alpha_1}{r^2} + \alpha_2 \frac{\log r}{r^2} - \lambda_0 \left( \frac{2M}{r} \right)^4,
\end{equation}

\begin{figure*}[htbp]
  \centering
  \includegraphics[width=\textwidth]{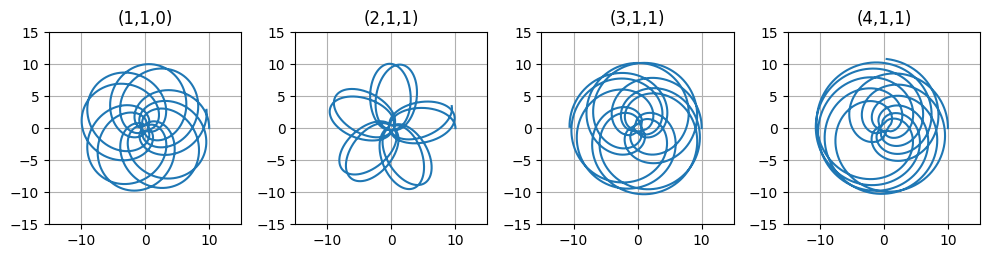}
  \includegraphics[width=\textwidth]{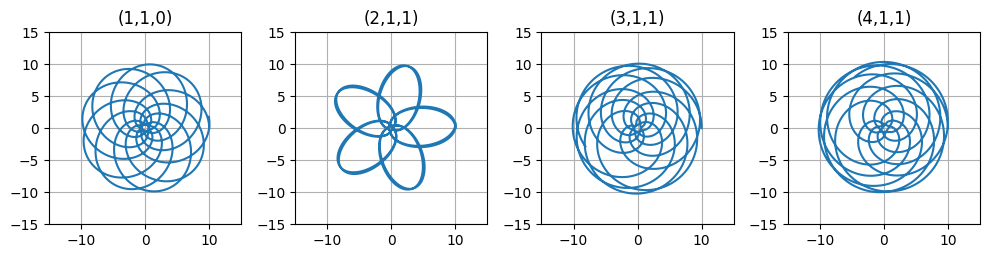}
  \caption{The Higher-Derivative Einstein-Æther theory solution's periodic orbits, c13 are set with distinct values in each column, namely 0, 0.00005, and 0.5.The images plot the (z, w, v) values for each orbit. The pictures in the upper row are for $\alpha\not=0, \beta\not=0$, and $\gamma\not=0$, whereas the pictures in the lower row are for $\alpha=0, \beta=0$, and $\gamma=0$.}
  \label{fig1}
\end{figure*}

where $\alpha_1$, $\alpha_2$, and $\lambda_0$ encode higher-order curvature and æther corrections:
\begin{align}
\alpha_1 &= -\frac{16\pi G_{ae}}{3} (3\alpha + \beta + 2\gamma) M, \\
\alpha_2 &= \frac{16\pi G_{ae}}{3} (\beta + 4\gamma) M, \\
\lambda_0 &= \frac{27 c_{13}}{256(1 - c_{13})}.
\end{align}

The particle’s geodesic motion in equatorial plane $\theta = \pi/2$ is governed by:
\begin{align}
\dot{t} &= \frac{E}{e(r)}, \quad \dot{\varphi} = \frac{L}{r^2}, \\
\dot{r}^2 &= E^2 - e(r) \left(1 + \frac{L^2}{r^2}\right).
\end{align}

Using $x = r \cos\varphi$, $y = r \sin\varphi$, and $z = 0$, we obtain:
\begin{align}
\ddot{x} &= \ddot{r} \cos\varphi - 2\dot{r}\dot{\varphi} \sin\varphi - r \dot{\varphi}^2 \cos\varphi, \\
\ddot{y} &= \ddot{r} \sin\varphi + 2\dot{r}\dot{\varphi} \cos\varphi - r \dot{\varphi}^2 \sin\varphi.
\end{align}

Substituting into the quadrupole derivatives, we get:
\begin{align}
h_+ &= \frac{G_{ae} m}{D_L} \left( \ddot{x}^2 - \ddot{y}^2 \right), \\
h_\times &= \frac{2 G_{ae} m}{D_L} \ddot{x} \ddot{y}.
\end{align}

In the adiabatic approximation (valid for EMRIs), energy and angular momentum losses are negligible over a few orbits. Thus, the waveform is computed from geodesic motion alone. For a given set of integers $(z, w, v)$, defining the orbit type (number of leaves, loops, orientation), one can solve the above system numerically to obtain $r(t)$ and $\varphi(t)$, and hence the waveforms.

 For an orbit with $(z,w,v) = (1,2,0)$, the particle completes one full angular cycle with two radial bounces, leading to a waveform with two oscillations per orbital period. The $h_+$ mode is maximal during periapsis, while $h_\times$ dominates when $\dot{r}\dot{\varphi}$ terms are strongest.

\begin{figure*}[htbp]
  \centering
  \includegraphics[width=\textwidth]{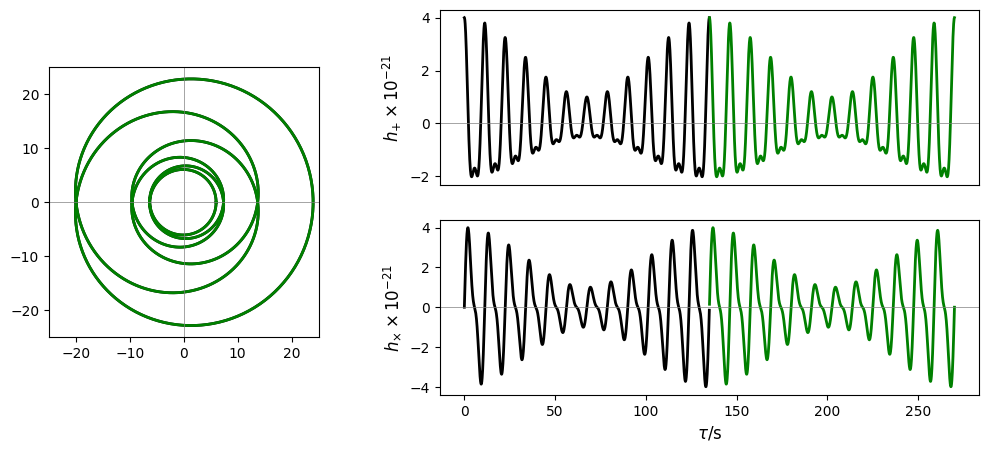}

  \caption{The left figure illustrates a particle moving from one apastron to another along a typical periodic orbit around a black hole, characterized by the parameters 
$(z,w,v)=(1,2,0)$. The right figure shows the modes of $h_+$ and $h_\times$ gravitational waves emitted by a black hole in Einstein-Æther theory with $q=1.5$.}
  \label{fig2}
\end{figure*}

Figures~\ref{fig1} and~\ref{fig2} show the periodic orbit and the corresponding gravitational waveform for two such examples. The impact of higher-derivative corrections is visible as phase shifts and distortions compared to general relativity, especially for $\gamma \neq 0$.

This methodology demonstrates how the zoom-whirl nature of relativistic periodic orbits maps directly to gravitational wave signatures, providing a potential channel to test higher-curvature and Lorentz-violating gravity in future space-based missions like LISA.

\pagebreak

\section{Discussion}

The direct detection of gravitational waves by the LIGO and Virgo collaborations  has transformed our ability to probe the strong-field, dynamical regime of gravity. While general relativity (GR) continues to pass observational tests with remarkable accuracy, numerous theoretical and phenomenological motivations suggest that GR may require modification at high energies. Effective field theory (EFT) approaches to quantum gravity predict that the Einstein-Hilbert action should be supplemented by higher-curvature corrections involving terms like $R^2$, $R_{\mu\nu} R^{\mu\nu}$, and $R_{\mu\nu\lambda\rho} R^{\mu\nu\lambda\rho}$ \cite{Stelle1977, Buchbinder1992, Woodard2007}. These terms naturally arise from the low-energy limit of string theory and loop quantum gravity and are expected to become relevant in the vicinity of compact astrophysical objects.

Simultaneously, theories that break local Lorentz invariance—such as Einstein-Æther theory provide a consistent framework in which the gravitational field couples to a dynamical, unit-norm timelike vector field. This theory modifies the structure of gravitational radiation and black hole solutions while maintaining diffeomorphism invariance. Linearized perturbation theory reveals up to five or six gravitational wave polarizations depending on the coupling constants . However, for a specific parameter space where $c_{14} = 0$, only the two tensor modes of GR propagate, making the theory compatible with existing gravitational wave observations.

Recently, there has been growing interest in how modified gravity theories alter the gravitational waveform emitted by compact binary systems \cite{Barausse2011, Blas2011, Stein2014}. Of particular importance are extreme-mass-ratio inspirals (EMRIs), where a stellar-mass object orbits a supermassive black hole, potentially undergoing thousands of orbits before merger. These systems offer a unique laboratory for testing deviations from GR through their intricate orbital dynamics and gravitational wave signatures.

In this work, we investigate the combined effects of higher-curvature corrections and æther field contributions on black hole spacetimes and gravitational radiation. We construct a perturbed static, spherically symmetric black hole solution in Einstein-Æther gravity including quadratic curvature terms. We analyze the structure of periodic orbits, with special focus on zoom-whirl trajectories, and compute the resulting gravitational waveforms using the quadrupole approximation. Our analysis shows that even minute deviations from GR can lead to measurable phase shifts and harmonic distortions in the waveform, potentially detectable with future space-based detectors such as LISA \cite{AmaroSeoane2017}.

This study highlights how the imprint of high-energy modifications to gravity may be encoded in the fine structure of gravitational wave signals and lays the groundwork for using waveform observations as probes of ultraviolet extensions of GR.

\section*{Acknowledgement}
 Research of GB is funded by the Science Committee of the Ministry of Science and Higher Education of the Republic of Kazakhstan (Grant AP19175860).

\bibliography{RefsGWB}

\providecommand{\href}[2]{#2}\begingroup\raggedright\begin{thebibliography}{10}

\bibitem{x}
S.~Lu and T.~Zhu, ``Gravitational radiations from periodic orbits around
  einstein-$\backslash$ae $\{$$\}$ ther black holes,'' {\em arXiv preprint
  arXiv:2505.00294} (2025) .

\bibitem{r1}
B.~P. Abbott, R.~Abbott, T.~D. Abbott, M.~R. Abernathy, F.~Acernese, K.~Ackley,
  C.~Adams, T.~Adams, P.~Addesso, R.~X. Adhikari, {\em et~al.}, ``Observation
  of gravitational waves from a binary black hole merger,'' {\em Physical
  review letters} {\bfseries 116} no.~6, (2016) 061102.

\bibitem{r2}
B.~P. Abbott, R.~Abbott, T.~Abbott, M.~Abernathy, F.~Acernese, K.~Ackley,
  C.~Adams, T.~Adams, P.~Addesso, R.~Adhikari, {\em et~al.}, ``Gw150914: First
  results from the search for binary black hole coalescence with advanced
  ligo,'' {\em Physical Review D} {\bfseries 93} no.~12, (2016) 122003.

\bibitem{r3}
B.~P. Abbott, R.~Abbott, T.~Abbott, M.~Abernathy, F.~Acernese, K.~Ackley,
  C.~Adams, T.~Adams, P.~Addesso, R.~Adhikari, {\em et~al.}, ``Properties of
  the binary black hole merger gw150914,'' {\em Physical review letters}
  {\bfseries 116} no.~24, (2016) 241102.

\bibitem{r4}
B.~P. Abbott, R.~Abbott, T.~Abbott, M.~Abernathy, F.~Acernese, K.~Ackley,
  C.~Adams, T.~Adams, P.~Addesso, R.~Adhikari, {\em et~al.}, ``Gw150914: The
  advanced ligo detectors in the era of first discoveries,'' {\em Physical
  review letters} {\bfseries 116} no.~13, (2016) 131103.

\bibitem{Stelle1977}
K.~S. Stelle, ``Renormalization of higher derivative quantum gravity,''
  \href{http://dx.doi.org/10.1103/PhysRevD.16.953}{{\em Phys. Rev. D}
  {\bfseries 16} (1977) 953}.

\bibitem{Buchbinder1992}
I.~L. Buchbinder, S.~D. Odintsov, and I.~L. Shapiro, {\em Effective Action in
  Quantum Gravity}.
\newblock IOP Publishing, 1992.

\bibitem{Woodard2007}
R.~P. Woodard, ``Avoiding dark energy with 1/r modifications of gravity,''
  \href{http://dx.doi.org/10.1007/978-3-540-71013-4_14}{{\em Lect. Notes Phys.}
  {\bfseries 720} (2007) 403}.

\bibitem{r5}
T.~Jacobson and D.~Mattingly, ``Gravity with a dynamical preferred frame,''
  {\em Physical Review D} {\bfseries 64} no.~2, (2001) 024028.

\bibitem{r6}
T.~Jacobson, ``Einstein-aether gravity: A status report,'' {\em arXiv preprint
  arXiv:0801.1547} (2008) .

\bibitem{r7}
C.~Eling, T.~Jacobson, and D.~Mattingly, ``Einstein-aether theory,'' in {\em
  Deserfest}, pp.~163--179.
\newblock World Scientific, 2006.

\bibitem{r7a}
T.~Jacobson and D.~Mattingly, ``Einstein-aether waves,'' {\em Physical Review
  D} {\bfseries 70} no.~2, (2004) 024003.

\bibitem{r8}
D.~Blas, O.~Pujolas, and S.~Sibiryakov, ``Models of non-relativistic quantum
  gravity: The good, the bad and the healthy,'' {\em Journal of High Energy
  Physics} {\bfseries 2011} no.~4, (2011) 1--53.

\bibitem{r9}
K.~Yagi and L.~C. Stein, ``Black hole based tests of general relativity,'' {\em
  Classical and Quantum Gravity} {\bfseries 33} no.~5, (2016) 054001.

\bibitem{r13}
J.~Levin and G.~Perez-Giz, ``A periodic table for black hole orbits,'' {\em
  Physical Review D—Particles, Fields, Gravitation, and Cosmology} {\bfseries
  77} no.~10, (2008) 103005.

\bibitem{r14}
J.~Levin and R.~Grossman, ``Dynamics of black hole pairs. i. periodic tables,''
  {\em Physical Review D—Particles, Fields, Gravitation, and Cosmology}
  {\bfseries 79} no.~4, (2009) 043016.

\bibitem{r15}
R.~Grossman and J.~Levin, ``Dynamics of black hole pairs. ii. spherical orbits
  and the homoclinic limit of zoom-whirliness,'' {\em Physical Review
  D—Particles, Fields, Gravitation, and Cosmology} {\bfseries 79} no.~4,
  (2009) 043017.

\bibitem{r16}
J.~Levin and G.~Perez-Giz, ``Homoclinic orbits around spinning black holes. i.
  exact solution for the kerr separatrix,'' {\em Physical Review D—Particles,
  Fields, Gravitation, and Cosmology} {\bfseries 79} no.~12, (2009) 124013.

\bibitem{r17}
G.~Perez-Giz and J.~Levin, ``Homoclinic orbits around spinning black holes. ii.
  the phase space portrait,'' {\em Physical Review D—Particles, Fields,
  Gravitation, and Cosmology} {\bfseries 79} no.~12, (2009) 124014.

\bibitem{r18}
J.~Healy, J.~Levin, and D.~Shoemaker, ``Zoom-whirl orbits in black hole
  binaries,'' {\em Physical review letters} {\bfseries 103} no.~13, (2009)
  131101.

\bibitem{r19}
J.~Levin, ``Energy level diagrams for black hole orbits,'' {\em Classical and
  Quantum Gravity} {\bfseries 26} no.~23, (2009) 235010.

\bibitem{r20}
V.~Misra and J.~Levin, ``Rational orbits around charged black holes,'' {\em
  Physical Review D—Particles, Fields, Gravitation, and Cosmology} {\bfseries
  82} no.~8, (2010) 083001.

\bibitem{r21}
R.~Grossman, J.~Levin, and G.~Perez-Giz, ``Harmonic structure of generic kerr
  orbits,'' {\em Physical Review D—Particles, Fields, Gravitation, and
  Cosmology} {\bfseries 85} no.~2, (2012) 023012.

\bibitem{r22}
K.~Glampedakis and D.~Kennefick, ``Zoom and whirl: Eccentric equatorial orbits
  around spinning black holes and their evolution under gravitational radiation
  reaction,'' {\em Physical Review D} {\bfseries 66} no.~4, (2002) 044002.

\bibitem{r23}
G.~Z. Babar, A.~Z. Babar, and Y.-K. Lim, ``Periodic orbits around a spherically
  symmetric naked singularity,'' {\em Physical Review D} {\bfseries 96} no.~8,
  (2017) 084052.

\bibitem{r24}
C.-Q. Liu, C.-K. Ding, and J.-L. Jing, ``Periodic orbits around kerr sen black
  holes,'' {\em Communications in Theoretical Physics} {\bfseries 71} no.~12,
  (2019) 1461.

\bibitem{r25}
M.~Azreg-A{\"\i}nou, Z.~Chen, B.~Deng, M.~Jamil, T.~Zhu, Q.~Wu, and Y.-K. Lim,
  ``Orbital mechanics and quasiperiodic oscillation resonances of black holes
  in einstein-{\ae}ther theory,'' {\em Physical Review D} {\bfseries 102}
  no.~4, (2020) 044028.

\bibitem{r26}
J.~Zhang and Y.~Xie, ``Probing a black-bounce-reissner--nordstr{\"o}m spacetime
  with precessing and periodic motion,'' {\em The European Physical Journal C}
  {\bfseries 82} no.~10, (2022) 854.

\bibitem{r27}
R.~Wang, F.~Gao, and H.~Chen, ``Periodic orbits around a static spherically
  symmetric black hole surrounded by quintessence,'' {\em Annals of Physics}
  {\bfseries 447} (2022) 169167.

\bibitem{r28}
Z.-Y. Tu, T.~Zhu, and A.~Wang, ``Periodic orbits and their gravitational wave
  radiations in a polymer black hole in loop quantum gravity,'' {\em Physical
  Review D} {\bfseries 108} no.~2, (2023) 024035.

\bibitem{r29}
X.-M. Deng, ``Periodic orbits around brane-world black holes,'' {\em The
  European Physical Journal C} {\bfseries 80} no.~6, (2020) 489.

\bibitem{r30}
X.-M. Deng, ``Geodesics and periodic orbits around quantum-corrected black
  holes,'' {\em Physics of the Dark Universe} {\bfseries 30} (2020) 100629.

\bibitem{r31}
H.-Y. Lin and X.-M. Deng, ``Bound orbits and epicyclic motions around
  renormalization group improved schwarzschild black holes,'' {\em Universe}
  {\bfseries 8} no.~5, (2022) 278.

\bibitem{r32}
Y.-Z. Li, X.-M. Kuang, and Y.~Sang, ``Precessing and periodic timelike orbits
  and their potential applications in einsteinian cubic gravity,'' {\em The
  European Physical Journal C} {\bfseries 84} no.~5, (2024) 529.

\bibitem{Barausse2011}
T.~J. E.~Barausse and T.~P. Sotiriou, ``Black holes in einstein-Æther and
  hořava–lifshitz gravity,''
  \href{http://dx.doi.org/10.1103/PhysRevD.83.124043}{{\em Phys. Rev. D}
  {\bfseries 83} (2011) 124043}.

\bibitem{Blas2011}
D.~Blas and E.~Lim, ``Phenomenology of theories of gravity without lorentz
  invariance: The preferred frame case,''
  \href{http://dx.doi.org/10.1142/S0218271814430091}{{\em Int. J. Mod. Phys. D}
  {\bfseries 23} (2014) 1443009}.

\bibitem{Stein2014}
L.~C. Stein and K.~Yagi, ``Equivalence principle and gravitational waves from
  extreme mass-ratio inspirals,''
  \href{http://dx.doi.org/10.1103/PhysRevD.89.044026}{{\em Phys. Rev. D}
  {\bfseries 89} (2014) 044026}.

\bibitem{AmaroSeoane2017}
P.~A.-S. et~al., ``Laser interferometer space antenna,'' {\em arXiv:1702.00786}
  (2017) .

\bibitem{r10}
N.~Yunes and X.~Siemens, ``Gravitational-wave tests of general relativity with
  ground-based detectors and pulsar-timing arrays,'' {\em Living Reviews in
  Relativity} {\bfseries 16} no.~1, (2013) 1--124.

\bibitem{r11}
E.~Berti, E.~Barausse, V.~Cardoso, L.~Gualtieri, P.~Pani, U.~Sperhake, L.~C.
  Stein, N.~Wex, K.~Yagi, T.~Baker, {\em et~al.}, ``Testing general relativity
  with present and future astrophysical observations,'' {\em Classical and
  Quantum Gravity} {\bfseries 32} no.~24, (2015) 243001.

\bibitem{r12}
S.~Mirshekari, N.~Yunes, and C.~M. Will, ``Constraining generic lorentz
  violation and the speed of the graviton with gravitational waves,'' {\em
  Phys. Rev. D} {\bfseries 85} no.~024041, (2012) 1110--2720.

\end{thebibliography}\endgroup
\bibliographystyle{utphys}

\end{document}